\documentclass[aps,prd,preprint,groupedaddress]{revtex4}
\usepackage{epsfig}
\usepackage{epstopdf}
\usepackage{latexsym}
\usepackage{graphicx}
\usepackage{dcolumn}
\usepackage{bm}

\newcommand{\be}{\begin{equation}}
\newcommand{\ee}{\end{equation}}
\newcommand{\bea}{\begin{eqnarray}}
\newcommand{\eea}{\end{eqnarray}}
\newcommand{\bi}{\begin{itemize}}
\newcommand{\ei}{\end{itemize}}
\def\vtr#1{{\bf #1}}
\def\mtx#1{{\bf #1}}

\def\flo#1{ #1} 

\begin{document}

\title{Matched filter for multi-transducers resonant GW antennas}
\author{Maria Alice Gasparini, Florian Dubath}
\email{alice.gasparini@physics.unige.ch, florian.dubath@physics.unige.ch}
\affiliation{D\'epartement de Physique Th\'eorique, Universit\'e de Gen\`eve,
24 quai Ernest-Ansermet, CH-1211 Gen\`eve 4\\}

\date{\today}

\begin{abstract}
We analyze two kinds of matched filters for data output of a spherical resonant GW detector. In order to filter the data of a real sphere, a strategy is proposed, firstly using an omnidirectional in-line filter, which is supposed to select periodograms with excitations, secondly by performing a directional filter on such selected periodograms, finding the wave arrival time, direction and polarization. We point out that, as the analytical simplifications occurring in the ideal 6 transducers TIGA sphere do not hold for a real sphere, using a 5 transducers configuration could be a more convenient choice.
\end{abstract}
\maketitle
\section{Introduction}

Spherical resonant gravitational wave (GW) detectors are the evolution of bar resonators. For an equivalent size they have a higher-cross section than other geometries and are the only kind of GW detectors able to to determine the signal arrival direction and polarization, even operated at a single location on Earth. Nowadays two spheres are under development in the Netherlands (MiniGRAIL)\cite{MG} and in Brasil (MARIO SCHENBERG)\cite{MS}.
Their experimental situation at present requires an efficient and concrete data-analysis strategy, \flo{which supplement the available literature (see for exemple \cite{GT},\cite{Merkowitz:1997qs})}. Works in this direction have already been done, see for example \cite{Caesar}.

In this paper we will focus on the case of matched filtering for such a detector, starting from some basic notions of \cite{Stevenson}. Matched filtering has been shown to be a key point in the burst oriented data analysis for resonant detectors \cite{[PIA]} and for interferometers~\cite{Schutz}.
We will discuss the question from a theoretical point of view, keeping in mind the technical implementation of our propositions. We propose a strategy in order to filter the sphere data, firstly by an omnidirectional in-line filter, which we call ``refined energy matched-filter", in order to select signal excitations, then by performing a ``directional matched-filter" on events selected by the first filter, extracting the wave arrival time, direction and polarization.

In actual experiments, the preferred configuration for the transducer placement is the truncated ichosaedral (TI) proposed by Merkowitz and Johnson \cite{Merkowity95}.
We find that when data analysis is performed for a realistic sphere with resonators and electronic readout, the analytical simplifications of an ideal sphere with transducers placed in the TI configuration do not hold anymore.  
Therefore using a 5 transducer configuration as the pentagonal hexacontahedron (PHC) proposed by Lobo and Serrano \cite{Lobo} can be a more convenient possibility, because simpler to build and manage.

This paper is structured as follows: in the next section we introduce notations and a schematic model of spherical detector. In section \ref{SF1} and \ref{SF2} we present two different ways of implementing a matched filter with a description of advantages and limitations for each one. In the last section we draw a strategy in order to perform data analysis with these two filters and end with our main conclusions.

\section{Model of spherical GW detector}
\subsection{\flo{Analytical considerations}}
We will provide a general frame to model a GW spherical detector equipped with mechanical resonators and electronic readouts. We only take into account couplings which are linear in each of the considered variables.

In accordance with previous literature, we call $\vtr{z}$ the amplitudes of the five fundamental normal modes ($\vtr{z}$ is then a five-component vector) and $\vtr{q}$ the positions of the six mechanical resonators in the TI configuration~\cite{Merkowity95,Merkowity95_2}. The model is completed by the current readouts $\vtr{I}$, which is a vector of $k\cdot6$ components, with $k=2$ for a system of capacitive single-SQUID transducers~\cite{FJL}.

In Fourier space the equations of motion for all those variables take the matrix form \cite{MichelsonZou,FJL}
\be
\mtx{Z}\pmatrix{\tilde{\vtr{z}}\cr\tilde{\vtr{q}}\cr\tilde{\vtr{I}}\cr}=\mtx{A} \pmatrix{\tilde{\vtr{F}}_S+\tilde{\vtr{F}}_{GW}\cr\tilde{\vtr{f}}_R\cr\tilde{\vtr{V}}}
\ee
where $\tilde{\vtr{F}}_S$ and $\tilde{\vtr{f}}_R$ represent the noise forces acting on the sphere and the resonators, $\tilde{\vtr{V}}$ the noise voltages and $\tilde{\vtr{F}}_{GW}$ the GW effect.
We use this equation in the form
\be
\pmatrix{\tilde{\vtr{z}}\cr\tilde{\vtr{q}}\cr\tilde{\vtr{I}}\cr}=\underbrace{\mtx{Z}^{-1}\mtx{A}}_{\mtx{G}} \pmatrix{\tilde{\vtr{F}}_S+\tilde{\vtr{F}}_{GW}\cr\tilde{\vtr{f}}_R\cr\tilde{\vtr{V}}}.
\ee
We have work with transducers in the TI configuration, the most used in recent literature, but this is not a unique choice. In this special case 
one can indeed easily find out the so called ``mode-channels" through the formula~\cite{Merkowity95,Merkowity95_2}
\be
\vtr{p}=\mtx{B}\vtr{I}_{\rm out} \label{mc},
\ee
$\vtr{I}_{\rm out}$ being the 6-vector containing the last six components of $\vtr{I},$ and $\mtx{B}$ the $5\times 6$ matrix representing the transducer positions. Its explicit expression can be found, for example, in \cite{SphereAlice}.
In the following we will always work with the output currents and therefore drop the subscript ``out''.
\subsection{\flo{Numerical model}}
The numerical model used in this paper is a direct sequel of those developed in \cite{FJL}. In the following we will restrict ourself to a 1 meter radius CuAl filled sphere, with a quality factor for the sphere's fundamental modes $Q_S=5\cdot 10^{7}$ with 6 capacitive transducers having both mechanical and electrical modes tuned on the sphere mode. We have tuned the transducers to the quantum  limit ($N_{\rm phonon}\equiv1$) rather than to use the present best sensitivity value ($N_{\rm phonon}\sim 50$) but this does not affect our analysis.

\section{Directional matched filter (DMF)\label{SF1}}

In a noise-free detector the output response to a GW burst is given in Fourier space by~\cite{MichelsonZou}
\be\label{e1}
\tilde{\vtr{I}}=-1/2\omega^2m_sR_s\chi\hat\mtx{G}\mtx{T}_V(\theta,\phi,\psi)\pmatrix{\tilde{h}_+\cr\tilde{h}_\times} 
\ee
where $ m_s,R_s\chi $ are the mass and the effective length of the sphere, $\hat\mtx{G}$ the $6\times5$ sub-matrix of $\mtx{G}$ relating the forces onto the modes to the output currents and $\mtx{T}_V(\theta,\phi,\psi)\pmatrix{\tilde{h}_+\cr\tilde{h}_\times}$ the projection of the GW tensor on the different modes, which depends on the wave polarization, $\psi$, and arrival direction, $(\theta,\phi).$ Choosing a coordinate system with origin on the center of the detector and the $z$ axis pointing the zenith, $\theta$ is defined to be the angle between the $z$ axis and the wave direction and $\phi$ the angle between its projection on the $xy$ plane and the $x$ axis. Using the tensor spherical harmonics formalism, which allows a definition for the five fundamental modes~\cite{MichelsonZou},  $\mtx{T}_V$ can be written as
\be
\mtx{T}_V=\pmatrix{\frac{\sqrt{3}}{2}\sin^2 \theta &0 \cr -\frac{1}{2}\sin 2\theta\sin\phi&\sin\theta\cos\phi\cr  \frac{1}{2}\sin 2\theta\cos\phi&\sin\theta\sin\phi       \cr                \frac{1}{2}\left(1-\cos^2\theta\right)\cos2 \phi&\cos\theta\sin2\phi\cr                -\frac{1}{2}\left(1-\cos^2\theta\right)\sin2 \phi&\cos\theta\cos2\phi}\pmatrix{\cos 2\psi&\sin 2 \psi\cr -\sin 2 \psi &\cos 2 \psi}.\label{TV}
\ee

We can then perform the matched filter searching the expected pattern, say $\tilde{\vtr{I}}_{\rm fil},$ into $\tilde{\vtr{I}}$, the output of the detector. This is performed using the noise weighted inner product~\cite{Stevenson} 
\be
\left( \tilde{\vtr{I}}_1\vert \tilde{\vtr{I}}_2\right)=\int\frac{d\omega}{2\pi}\tilde{\vtr{I}}_1^\dagger(\omega)\mtx{S}^{-1}(\omega)\tilde{\vtr{I}}_2(\omega)
\ee
where $\mtx{S}$, the noise matrix, is the expectation value of the product $\vtr{I}_{\rm noise}\vtr{I}_{\rm noise}^\dagger$ with $\vtr{I}_{\rm noise}$ representing the output due to noises only~\cite{Stevenson}. For experimental purposes, the noise matrix can be computed as an average over the detector output $\mtx{S}=\langle\vtr{I}\vtr{I}^\dagger\rangle$. The output due to noises is given by
\be
\vtr{I}_{\rm noise}=\overline{\mtx{G}} \pmatrix{\tilde{\vtr{F}}_S\cr\tilde{\vtr{f}}_R\cr\tilde{\vtr{V}}}
\ee
with $\overline\mtx{G}$ being the sub-matrix given by the last $6$ lines of $\mtx{G}$. In order to rewrite the matrix $\mtx{S}$ in terms of $\overline\mtx{G}$ and the noise spectral densities, we use the statistical properties and the independence of  the noises $\tilde{\vtr{F}}_S$ $ \tilde{\vtr{f}}_R$ and $\tilde{\vtr{V}}$ ~\cite{FJL}.

We obtain the filtered signal as
\be
\gamma(\omega)=\frac{\tilde{\vtr{I}}_{\rm fil}^\dagger(\omega)\mtx{S}^{-1}(\omega)\tilde{\vtr{I}}(\omega)}{\left( \tilde{\vtr{I}}_{\rm fil}\vert \tilde{\vtr{I}}_{\rm fil}\right)}.
\ee
${\vtr{I}}_{\rm fil}$ is the filter output obtained by choosing a waveform and a polarization of $h_+^{fil}(t)\, ,h^{fil}_\times(t)$ and a direction for the expected wave.
The coefficient of the projection on the pattern is
\be
\Gamma=\int\frac{d\omega}{2\pi}\gamma(\omega)=\frac{\left( \tilde{\vtr{I}}_{\rm fil}\vert \tilde{\vtr{I}}\right)}{\left( \tilde{\vtr{I}}_{\rm fil}\vert \tilde{\vtr{I}}_{\rm fil}\right)}.
\ee

The same equations hold using mode-channels: one has just to multiply eq.(\ref{e1}) on the left by $\mtx{B}$ and perform the replacement $\vtr{I}\rightarrow\vtr{p}.$

\subsection{Limitations}

The principal problem with this kind of matched filter is that $\tilde{\vtr{p}}_{\rm fil}$ is a function of the arrival direction and of the polarization angle. Therefore such a filter can be built for a specific GW  direction, polarization and shape (usually matched filters are built for $\delta$-like excitations \flo{and we will perform our test with such a waveform} but one can also use sin-Gaussian or a specific template) which therefore are supposed to be known ``a priori". This is not a realistic possibility.  Such a filter is unable to detect with high efficiency an excitation with a different arrival direction or polarization. 

\begin{figure}
 \begin{center}
\includegraphics[width=8cm]{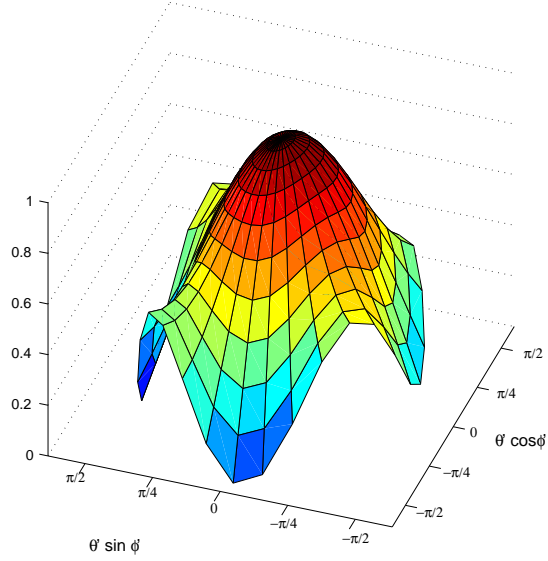} 
\caption{\footnotesize Direction dependence of $s$ on the cylindrical coordinates $\theta',\,\phi'$. $\theta'$ is the radial coordinate, and $\phi'$ the angular coordinate. We have fixed $\theta=0$, $\phi=0$,$\psi=\pi/4$, $\tilde h^{\rm fil}_+=1$, $\tilde h^{\rm fil}_\times=0$ and plot $s$ for $\theta'\in[0,\pi/2]$ and $\phi'\in[0,2\pi]$. In order to avoid a mulitvalued function at $\theta'=0$ we have corrected the polarization angle setting $\psi'=\psi+\phi'$. We use the sphere model with capacitive transducers described in~\cite{FJL}.  \label{f1}}
 \end{center}
 \end{figure}

The loss of sensitivity of the directional filter constructed for the direction $\theta,\,\phi$ and polarization $\psi$ when a GW is coming from another direction, say $\theta'$ and $\phi'$, with another polarization, $\psi'$, is obtained by fixing $\tilde\vtr{h}$ and computing the ratio
\be
s=\frac{\left( \tilde{\vtr{I}}_{\rm fil}(\theta,\phi,\psi)\vert \tilde{\vtr{I}}_{\rm fil}(\theta',\phi',\psi')\right)}{\left( \tilde{\vtr{I}}_{\rm fil}(\theta,\phi,\psi)\vert \tilde{\vtr{I}}_{\rm fil}(\theta,\phi,\psi)\right)}\ .
\ee
In Figure \ref{f1}, we plot the $\theta'$ and $\phi'$ dependence of $s$ for \flo{our sphere model}.

The polarization dependence of $s$, for a given direction ($\theta=\theta'$, $\phi=\phi'$),  can be computed using the explicit $\psi$ dependence of eq.(\ref{TV}). Without loss of generality we can fix $\tilde h^{\rm fil}_+=1$, $\tilde h^{\rm fil}_\times=0$ and $\psi'=\psi+\delta\psi$, getting
\bea
s&=&\pmatrix{1&0}\pmatrix{\cos 2\psi&\sin 2 \psi\cr -\sin 2 \psi &\cos 2 \psi}^\dagger \mtx{M}\pmatrix{\cos (2\psi')&\sin  (2\psi')\cr -\sin  (2\psi') &\cos (2\psi') }\pmatrix{1\cr0}\nonumber\\
&=&\frac{1}{2}\left(\sin(4\psi+2\delta\psi)+\sin(2\delta\psi)\right)\mtx{M}_{11}+\nonumber\\
&&\hspace{2cm} \frac{1}{2}\left(-\cos(4\psi+2\delta\psi)+ \cos(2\delta\psi)\right)\mtx{M}_{22}.
\eea
All terms which are independent of $\psi$ are put into the $2\times 2$ matrix $\mtx{M}$. As an exemple, for $\psi=\pi/4$ the sensitivity of the filter drops as $\cos(2\delta\psi)$. For each direction, reconstructing the polarization is possible by fixing a polarization angle $\psi$ and performing two identical filters up to a shift of $\pi/4$ in the polarization angle. \\
If the wave arrival direction is not known a priori (as it is always the case), one possibility is to produce several filters for different directions. The detector output has then to be processed with all these filters, giving finally a direction and a polarization which maximize $\Gamma$. The clear advantage is that this procedure gives an estimate of both the arrival direction (the one corresponding to the ``better" filter) and the polarization. However this procedure needs high computing power since it requires $6\times N$ more computational time than resonant bar matched filtering, $N$ being the number of filters. We can estimate $N$ using Figure $\ref{f1}$. If we want an efficiency of say $s\geq0.8,$ we can deduce that a filter with arrival direction in $\theta=0$ can only see signals from directions with $\theta\leq\pi/6$ and therefore cover $\sim1/8$ of the hemisphere (remember that antipodal directions are identified). Taking into account the polarization, we obtain that $N$ has to be at least of order $20$. However, once a burst has been caught by the filter, a hierarchical algorithm has to set up  (trying maybe also some other wave forms) in order to reconstruct an accurate arrival direction. This is not satisfactory for a in-line analysis but can be useful in order to analyze particularly some interesting events. 

\subsection{Performance}

\begin{figure}
 \begin{center}
\includegraphics[width=8cm]{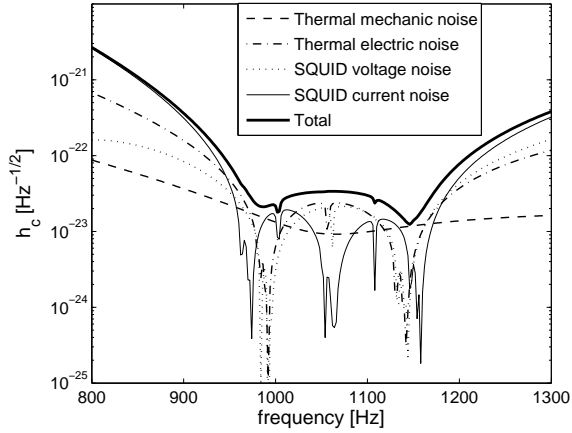} 
\caption{\footnotesize Strain sensitivity computed for directional filter. We use the 1[m] radius sphere model with capacitive transducers of~\cite{FJL} and display the different noise contributions. We have used a $\delta$-filter \label{fHcD}}
 \end{center}
 \end{figure}

The sensitivity of a detector depends on the method used to analyze its data. We quantify now the sensitivity that can be reached with the directional matched filter by computing its ``strain sensitivity", defined as the amplitude $\tilde{h}_c(\omega)$ that an incoming GW needs in order to produce the same output as detector internal noise, that is to have a signal to noise ratio (SNR ) of 1.  As expected, for matched filtering the highest SNR is reached when the filter has the same form as the signal (when $\tilde{\vtr{I}}$ is equal to $\tilde{\vtr{I}}_{\rm fil}$), see~\cite{Stevenson} and references therein. This situation corresponds to  optimal filtering.  In this case we can define the filter SNR  density as
\be \label{SNR}
\sigma(\omega)={\tilde{\vtr{I}}_{\rm fil}}^{+\dagger}(\omega)\mtx{S}^{-1}(\omega)\tilde{\vtr{I}}_{\rm fil}^+(\omega)\ .
\ee
Furthermore, without loss of generality, we can perform a rotation of the polarization frame in order to set one of the polarizations equal to~$0,$
\bea
\psi&\rightarrow&\psi'\\
\tilde{h}^{\rm fil}_+&\rightarrow&\tilde{h}^{\rm fil}\\
\tilde{h}^{\rm fil}_\times&\rightarrow&0
\eea
Using these new variables together with eq.~(\ref{e1}), we get
\be
\sigma(\omega)=f(\omega)\vert\tilde{h}^{\rm fil}(\omega)\vert^2
\ee 
where $f(\omega)$ is a proportionality coefficient.
Setting $\sigma(\omega)\equiv1$ gives the strain sensitivity 
\be
\tilde{h}_c(\omega)=\sqrt{\frac{1}{f(\omega)}}\ .
\ee
In Figure~\ref{fHcD}, $\tilde{h}_c(\omega)$ is plotted for a directional matched filter applied to a 1[m] radius sphere with six single SQUID capacitive transducers in TI configuration as described in \cite{FJL}. We stress that such a strain sensitivity is reached only for optimal filtering, and it will be reduced if the filter is not built with exactly the same arrival direction and waveform as the received GW signal.

\section{Energy matched filtering\label{SF2}}
We observe that the sum of the squared excitations of the five fundamental sphere modes 
\be
\pmatrix{\tilde{h}^*_+&\tilde{h}^*_\times}\mtx{T}_V^\dagger\mtx{T}_V\pmatrix{\tilde{h}_+\cr\tilde{h}_\times}, 
\ee  
 is proportional to $\vert\tilde{h}_+\vert^2+\vert\tilde{h}_\times\vert^2$ and independent of the direction. 
Furthermore, one of the most useful properties of the mode-channels formalism is that the vector $\vtr{p}$ is proportional to the fundamental mode excitations of the sphere, at least when considering only an ideal sphere with TI resonators. In other words, this means that $\mtx{B}\hat\mtx{G}$ is proportional to the identity. The proportionality factor can indeed be easily worked out ~\cite{SphereAlice}.

As a consequence $\hat\mtx{G}^\dagger\mtx{B}^\dagger\mtx{B}\hat\mtx{G}$ is proportional to the identity too, and the norm of the vector $\vtr{p}$ in Fourier space 
\bea\label{e2}
\tilde{E}_{\rm fil}(\omega)&=&\tilde{\vtr{p}}_{\rm fil}^\dagger\tilde{\vtr{p}}_{\rm fil}=\tilde{\vtr{I}}_{\rm fil}^\dagger\mtx{B}^\dagger\mtx{B}\tilde{\vtr{I}}_{\rm fil}\nonumber\\
&=&1/4\omega^4m_s^2R_s^2\chi^2 \pmatrix{\tilde{h}^{*\rm fil}_+&\tilde{h}^{*\rm fil}_\times}\mtx{T}_V^\dagger \hat\mtx{G}^\dagger\mtx{B}^\dagger\mtx{B}\hat\mtx{G}\mtx{T}_V\pmatrix{\tilde{h}^{\rm fil}_+\cr\tilde{h}^{\rm fil}_\times}\nonumber\\
\eea
has the same kind of proportionality to $\vert\tilde{h}^{\rm fil}_+\vert^2+\vert\tilde{h}^{\rm fil}_\times\vert^2$ and directional independence\footnote{Note that, if in the TI configuration $\mtx{B}\mtx{B}^\dagger$ is proportional to the identity, this is not the case for $\mtx{B}^\dagger\mtx{B}$.}.

The main idea of this section is to use this directional independence
in order to perform a matched filtering on the total energy stored into the sphere, which can be computed from the detector signal \be
\tilde{E}(\omega)=\tilde{\vtr{p}}^\dagger\tilde{\vtr{p}}=\tilde{\vtr{I}}^\dagger\mtx{B}^\dagger\mtx{B}\tilde{\vtr{I}}\ .
\ee 
The noise weighted inner product is defined as 
\be
\left( \tilde{E}_1\vert \tilde{E}_2\right) =\int\frac{d\omega}{2\pi}\frac{\tilde{E}_{1}(\omega)\tilde{E}_{2}(\omega)}{\tilde{E}_{\rm noise}^2(\omega)}
\ee
where the noise contribution 
 \be
\tilde{E}_{\rm noise}=\tilde{\vtr{I}}_{\rm noise}^\dagger\mtx{B}^\dagger \mtx{B}\tilde{\vtr{I}}_{\rm noise}
\ee
can also be computed by means of the noise spectral densities.\\
For a given signal $E_{\rm fil}$, the filter gain is then
\be
R=\left( \tilde{E}_{\rm fil}\vert \tilde{E}_{\rm fil}\right) \label{R}
\ee
and the filter is given by
\be
   \gamma_E(\omega)=\frac{1}{R}\frac{\tilde{E}_{\rm fil}(\omega)\tilde{E}(\omega)}{\tilde{E}_{\rm noise}^2(\omega)}.
\ee
Integrating we get
\be
\Gamma_E=\int\frac{d\omega}{2\pi}  \gamma_E(\omega)=\frac{\left( \tilde{E}_{\rm fil}\vert \tilde{E}\right) }{\left( \tilde{E}_{\rm fil}\vert \tilde{E}_{\rm fil}\right) }. 
\ee

We define the efficiency with respect to the directions and polarizations in the same way as we have done for the directional filter:
\be
s_E=\frac{\left( \tilde{E}_{\rm fil}(\theta,\phi,\psi)\vert \tilde{E}_{\rm fil}(\theta',\phi',\psi')\right) }{\left( \tilde{E}_{\rm fil}(\theta,\phi,\psi)\vert \tilde{E}_{\rm fil}(\theta,\phi,\psi)\right) }. 
\ee
Since we have defined the energy matched filter to avoid angular dependence, we expect that $s_E\equiv1$. 

Once the filter waveform  has been fixed, its amplitude is determined by a single parameter $(h^{\rm fil })^2$. Typically, for a $\delta$-like excitation we have $\vert\tilde{h}^{\rm fil}_+(\omega)\vert^2+\vert\tilde{h}^{\rm fil}_\times(\omega)\vert^2=(h^{\rm fil})^2=\rm constant$ and for a sin-Gaussian $\vert\tilde{h}^{\rm fil}_+(t)\vert^2+\vert\tilde{h}^{\rm fil}_\times(t)\vert^2=(h^{\rm fil})^2\sin^2(\omega_0t)e^{-2\gamma(t-t_0)^2}$. We can build the filter $\tilde{E}_{\rm fil}(\omega)$ from eq.~(\ref{e2}) with the chosen pattern and $h^{\rm fil}=1$. In this case, computing $\Gamma_E$ on a detector output periodogram $\tilde{E}$ returns the amplitude of an eventual hosted excitation having the same waveform as the filter:
\be
\Gamma_E =h^2\label{e3}.
\ee

\subsection{Limitations}
The energy matched filter is efficient to determine the presence of an excitation out of noise, independently on the direction and polarization of the wave. Consequently it cannot be used in order to determine the arrival direction of the GW but rather as a in-line filter able to detect efficiently the presence of a GW signal.  However, this supposes that $\mtx{B}\hat\mtx{G}$ is proportional to the identity, which is not true for a real spherical detector. For a physical model of resonant spheres, we have to include the splitting in frequency due to the Earth gravity field and electronic readout. 

It is interesting to \flo{split} $\mtx{B}\hat\mtx{G}$ in order to quantify the deviation to a multiple of the identity
\be
 \mtx{N}(\omega)=\mtx{B}\hat\mtx{G}(\omega)-C(\omega)\mtx{1}_5, \ \ C(\omega)=\frac{1}{5}\sum_{i=1}^{5}\left(\mtx{B}\hat\mtx{G}(\omega)\right)_{ii}
\ee
where $\mtx{1}_5$ is the $5\times5$ identity matrix. 

\begin{figure}
 \begin{center}
\includegraphics[width=8cm]{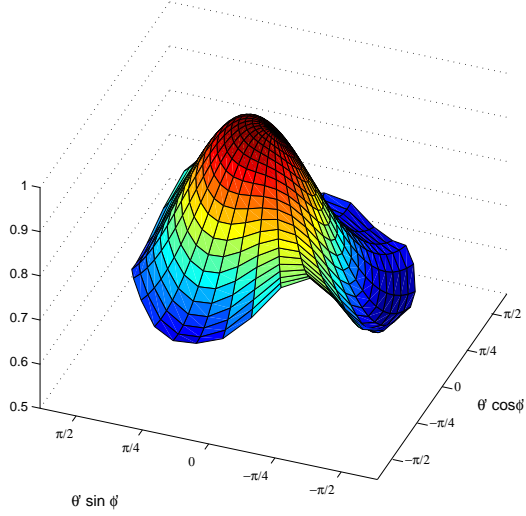} 
\caption{\footnotesize Under the same conditions than Figure \ref{f1} we plot the direction dependence of $\sqrt{s_E}$. Note that $\sqrt{s_E}$ vary only between $\sim 1/2$ and 1. \label{f2}}
 \end{center}
 \end{figure}

Numerical investigations for models with capacity transducers, show that the elements of $\mtx{N}$ may be non negligible against $C$. In particular for our model the magnitude of the biggest elements of the matrix $\mtx{N}$ is typically of the order of 7\% of $C$, but can reach 400\% of $C$ for $\omega$ near the resonances. As a sequel, $\tilde{E}(\omega)$ is direction dependent, as shown in Figure \ref{f2}.

\subsection{Refined energy matched filter (REMF)}
We adapt our analysis in order to recover the direction-independence of $\tilde{E}(\omega)$. With this aim, we change the mode-channel definition (\ref{mc}). We need to replace the matrix $\mtx{B}$ by a matrix $\hat{\mtx{B}}(\omega)$ such that the product $\hat{\mtx{B}}(\omega)\hat\mtx{G}(\omega)$ is proportional to the identity.
We define the matrix
\be
    \hat{\mtx{B}}=\mtx{b}\pmatrix{\mtx{1}_5&\pmatrix{1\cr1\cr1\cr1\cr1}}.
\ee
The choice of the last column is arbitrary. For each value of $\omega$ we can compute the $5\times5$ matrix $\mtx{b}$ as
\be
\mtx{b}=\left[ \pmatrix{\mtx{1}_5&\pmatrix{1\cr1\cr1\cr1\cr1}}\hat{\mtx{G}}(\omega)\right]^{-1}
\ee
and the new mode-channels $\tilde{\vtr{p}}(\omega)$ are defined as 
\be
\tilde{\vtr{p}}(\omega)=\hat{\mtx{B}}(\omega)\tilde{\vtr{I}} (\omega).
\ee

Note that the matched filter does work since we fix a unique set of matrices $\hat{\mtx{B}}(\omega)$ replacing the matrix $\mtx{B}$ in equations from (\ref{e2}) to (\ref{e3}).

The main consequences of this change are the following:
\begin{itemize}
\item We need the knowledge of $\hat{\mtx{G}}(\omega)$ in order to compute the mode-channels.
\item The mode-channels are only defined through the Fourier space computation.
\item The product $\hat\mtx{B}(\omega)\hat\mtx{G}(\omega)$ is proportional to the identity but the proportionality constant is now $C(\omega)$, an arbitrary (but non zero) function of $\omega$.
\item $\gamma_E(\omega)$ has to be manipulated with some caution since it depends on $C(\omega)$. 
However the property $\Gamma_E=h^2 $ remains true and is sufficient in order to select periodograms hosting excitations. 
\end{itemize}

\subsection{Performance}

We checked with our numerical model \flo{and for $\delta$-like waveform} that the redefined matched energy filter is able to give a response which is independent of the polarization and the direction (of both the filter function and the incoming wave) with relative error compatible with numerical errors $O(10^{-16})$.

\begin{figure}
 \begin{center}
\includegraphics[width=8cm]{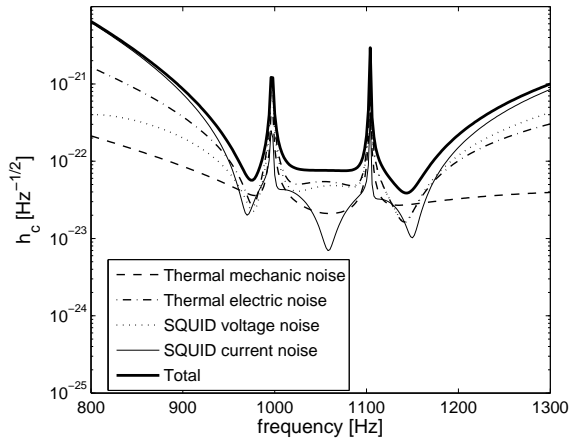} 
\caption{\footnotesize Strain sensitivity obtained for the energy filter. This computation is performed for the same sphere model as in Fig.~\ref{fHcD}.  \label{fHcE}}
 \end{center}
 \end{figure}

We expect the strain sensitivity corresponding to REMF to be worse than the one obtained from DMF. In particular, the computation of $E$ as a sum of the {\it squared} mode-channels, implies that we loose the relative phase and therefore the possibility of performing a coherent analysis~\cite{FJL}. In order to compute the strain sensitivity for the REMF, we need the SNR density $\sigma_E(\omega)$, which is given by the integrated term in eq.~(\ref{R}). This is, by construction, proportional to $\left(\vert\tilde{h}^{\rm fil}_+\vert^2+\vert\tilde{h}^{\rm fil}_\times\vert^2 \right)^2$ and the strain sensitivity is then obtained as the  4th-root of the proportionality coefficient. Figure~\ref{fHcE} shows the strain sensitivity of the refined energy matched filter. Note the presence of horns, typical of non-coherent analysis~\cite{FJL}.

We emphasize that even if the sensitivity is worse for the energy filter, this is \flo{partially} compensated by the fact that the energy filter performance is the same for all the directions and polarizations.

\section{Conclusions and future developments}

In this paper we have analyzed two ways for matched-filtering data output of a spherical GW detector. In section \ref{SF2} the ``refined energy matched filter" has been presented, whose efficiency does not depend on the GW direction.

For the specific setup of each existing GW spherical detector, it would be worth to investigate in more details the performance of the pipeline data analysis proposed in this work: firstly using the REMF as an in-line filter in order to select periodograms with excitations. REMF requires some computer memory, since the set of matrices $\hat\mtx{B}(\omega)$ has to be stored, but only few more stages of computation than the matched filter used for the cylindrical detectors.

Then, on periodograms selected by REMF, the directional matched filter presented in section \ref{SF1} of this paper can be used in order to extract the arrival time, direction and polarization of the GW signal. Furthermore, once we have obtained (for a chosen $\psi$) $\tilde{h}_+$ and  $\tilde{h}_\times$ from the DMF procedure, we have a check of the coherence of the whole procedure by making the comparison with the value of $\vert\tilde{h}_+\vert^2+\vert\tilde{h}_\times\vert^2 $ obtained from the refined energy matched filter. 

Finally we stress that the mode-channel simplifications occurring when the 6-transducers TI configuration is considered do not hold in a realistic model of a spherical antenna. Therefore, for the data analysis strategy proposed in this work, it may be convenient to choose working with only five transducers.  In such a case the $\hat\mtx{G}$ is a $5\times5$ matrix and, finding a configuration of the transducer such that it is invertible for all values of $\omega$, we can perform a REMF with $\hat{\mtx{B}}=\hat{\mtx{G}}^{-1}$.
\\

We would like to thank C. A. Costa and A. Malaspinas. This work is partially supported by the Swiss National Science Foundation (FNS).

\end{document}